\documentclass[11pt]{article}
\usepackage{appendix}

\usepackage{amssymb,amsmath}

\usepackage[figuresright]{rotating}

\usepackage[applemac]{inputenc}

\usepackage{tikz}
\usepackage{pgfplots}

\usepgfplotslibrary{groupplots} 
\usetikzlibrary{pgfplots.groupplots}

\usepackage{graphicx}
\usepackage{dcolumn}
\usepackage{bm}
\usepackage{amscd,amsthm}
\usepackage{ifthen}
\usepackage{booktabs}
\usepackage{xfrac}
\usepackage{bm}

\usepackage{textcase}

\usepackage{hyperref}
\hypersetup{
    bookmarks=true,         
    unicode=false,          
    pdftoolbar=true,        
    pdfmenubar=true,        
    pdffitwindow=false,     
    pdfstartview={FitH},    
    pdfauthor={Reinhard Heckel},     
    pdfsubject={Subject},   
    pdfcreator={Reinhard Heckel},   
    pdfproducer={Producer}, 
    pdfnewwindow=true,      
    colorlinks=true,       
    linkcolor=red,          
    citecolor=green,        
    filecolor=magenta,      
    urlcolor=cyan           
}

\newtheorem{theorem}{Theorem}

\newtheorem{proposition}{Proposition}

\definecolor{DarkBlue}{rgb}{0,0,0.7}

\newcommand{\K}{{K}}

\newcommand{\derd}{\partial}

\usepackage{rays_defs_13}

\newcommand{\setA}{\mathcal A}

\renewcommand{\L}{L}

\newcommand{\N}{N}
\renewcommand{\S}{S}

\newcommand{\FK}{F} 

\newcommand{\sign}{\mathrm{sign}}

\newcommand{\tauc}{\bar \tau}
\newcommand{\nuc}{\bar \nu}

\newcommand{\minlet}[1]{\tilde #1} 

\newcommand\SRF{\mathrm{SRF}}
\newcommand{\atom}{\vf}
\newcommand{\atomcen}{\tilde \vf}

\renewcommand\footnotemark{}
\title{Super-Resolution MIMO Radar}

\author{Reinhard Heckel
 \\[0.5em]
  \multicolumn{1}{p{.7\textwidth}}{\centering 
  Department of Electrical Engineering and Computer Sciences\\
UC Berkeley,  Berkeley, CA}
}

\usepackage[
backend=bibtex,
url=false,isbn=false,doi=false,
hyperref=auto,
style=alphabetic,
natbib=true,
sorting=nty,
firstinits=true,
maxbibnames=10,
]{biblatex}

\newbibmacro*{journal}{%
  \iffieldundef{journaltitle}
    {}
    {\printtext[journaltitle]{%
       \printfield[myplain]{journaltitle}%
       \setunit{\subtitlepunct}%
       \printfield[myplain]{journalsubtitle}}}}
\DeclareFieldFormat[article,inproceedings,incollection]{titlecase}{\MakeSentenceCase{#1}}

\bibliography{../bibliography.bib} 

\newcommand{\mAA}{\mA}

\begin{document}

\maketitle

\begin{abstract}
A multiple input, multiple output (MIMO) radar emits probings signals with multiple transmit antennas and records the reflections from targets with multiple receive antennas. 
Estimating the relative angles, delays, and Doppler shifts from the received signals allows to determine the locations and velocities of the targets. 
Standard approaches to MIMO radar based on digital matched filtering or compressed sensing only resolve the angle-delay-Doppler triplets on a $(1/(N_T N_R), 1/B,1/T)$ grid, where $N_T$ and $N_R$ are the number of transmit and receive antennas, $B$ is the bandwidth of the probing signals, and $T$ is the length of the time interval over which the reflections are observed. 
In this work, we show that the \emph{continuous} angle-delay-Doppler triplets and the corresponding attenuation factors can be recovered perfectly by  solving a convex optimization problem. 
This result holds provided that the angle-delay-Doppler triplets are separated either by $10/(N_T N_R-1)$ in angle, $10.01/B$ in delay, or $10.01/T$ in Doppler direction. 
Furthermore, this result is optimal (up to log factors) in the number of angle-delay-Doppler triplets that can be recovered. 
\end{abstract}

\section{Introduction}
Traditional pulse-Doppler radar systems transmit a probing signal and receive the reflections from the targets with a single antenna. 
By estimating the induced delays and Doppler shifts such a Single-Input Single-Output (SISO) radar can determine the relative distances and velocities of the targets. 
A SISO radar can, however, not determine the actual positions of the objects with a single measurement. 
MIMO radar systems \cite{bliss_multiple-input_2003,li_mimo_2007} use multiple antennas to transmit probing signals simultaneously and record the reflections from the targets with multiple receive antennas. 
A MIMO radar can thereby, in principle, resolve the relative angles along with the relative distances and velocities of targets with a single measurement. 
In this paper, we study the problem of recovering the \emph{continuous} angels, delays and Doppler shifts from the response to known and suitably selected probing signals. 
As we will see later, this problem---termed the super-resolution MIMO radar problem---amounts to recover a signal that is sparse in a \emph{continuous} dictionary from linear measurements. 

In case the targets may be assumed to lie on a sufficiently \emph{coarse} grid, 
compressed sensing \cite{candes_robust_2006} based  approaches provably recover the angle-delay-Doppler triplets for MIMO \cite{dorsch_refined_2015,strohmer_adventures_2015}, and the delay-Doppler pairs for SISO \cite{herman_high-resolution_2009,baraniuk_compressive_2007,heckel_identification_2013} radar. 
However, to establish those results, the papers~\cite{herman_high-resolution_2009,baraniuk_compressive_2007,dorsch_refined_2015,strohmer_adventures_2015} 
assume that angles, delays, and Doppler shifts lie on a sufficiently \emph{coarse} grid, namely a grid with spacing $1/(N_TN_R), 1/B$, and $1/T$, in angle, delay, and Doppler direction, respectively. 
Here, $N_T$ and $N_R$ are the number of transmit and receive antennas, $B$ is the bandwidth of the probing signals, and $T$ is the time interval over which the reflections are observed. 
Since $N_T,N_R, B$, and $T$ are 
physical problem parameters, they can in general not be made (arbitrarily) large in order to make the grid finer. 
The coarseness of the grid is needed for the measurement matrix to be sufficiently incoherent, therefore the aforementioned results cannot be extended to a grid with  significantly finer spacing. 
In some special cases, however, off the grid recovery is possible with standard spectral estimation techniques. 
For example, 
in case of a single input antenna and either known and constant delays, or known and constant Doppler shifts, 
the super-resolution radar problem reduces to a standard 2D line spectral estimation problem \cite[Sec.~5]{strohmer_adventures_2015}. 
 For those special cases, the target locations can be recovered---off the grid---with standard spectral estimation techniques such as Prony's method, MUSIC, and ESPRIT \cite{stoica_spectral_2005}. 
In general, however, the super-resolution MIMO radar problem cannot be reduced to a line spectral estimation problem, not even in the SISO case. Therefore, traditional spectral estimation techniques are not directly applicable. 
Recently, an alternative, convex optimization based approach to solve line spectral estimation and related 
problems has been proposed. 
Specifically, in~\cite{candes_towards_2014} it is shown that the corresponding frequencies can be recovered perfectly by solving a convex total-variation norm minimization program, provided they are sufficiently separated. 
Related convex programs have been studied for compressive sensing off the grid \cite{tang_compressed_2013}, denoising \cite{bhaskar_atomic_2012}, 
signal recovery from short-time Fourier measurements \cite{aubel_theory_2015}, 
and the SISO super-resolution radar problem \cite{heckel_super-resolution_2015}. 
The super-resolution MIMO radar problem, however, is more difficult than its SISO counterpart studied in \cite{heckel_super-resolution_2015}, due to the additional angle dimension, and since the probing signals from different transmit antennas superimpose at the receive antennas. 
 
In this work we propose a convex program similar to those in \cite{candes_towards_2014,tang_compressed_2013,bhaskar_atomic_2012,heckel_super-resolution_2015,chandrasekaran_convex_2012}, and show that it recovers the \emph{continuous} angles, delays, and Doppler shifts perfectly, provided that they are sufficiently separated. 
To the best of our knowledge, this is the first approach that \emph{provably} recovers the angle-delay-Doppler triplets \emph{off the grid} under general conditions. 
Furthermore, we show that a simple convex $\ell_1$-minimization program can recover the angles and delay-Doppler shifts on an \emph{arbitrarily fine} grid, again provided they are sufficiently separated. 
Finally, we provide numerical results demonstrating that our approach is robust to noise. 

\paragraph{Outline: }
The remainder of this paper is organized as follows. Section \ref{sec:model} contains the MIMO radar model and formal problem statement. 
In Sections~\ref{sec:atomicnmin} and \ref{sec:mainares} we present our convex optimization based recovery approach and corresponding performance guarantees. 
In Section~\ref{sec:discretesuperes} we show that $\ell_1$-minimization recovers the locations on an arbitrarily \emph{fine} grid, and in Section~\ref{sec:numres} we provide numerical results demonstrating that our approach is robust to noise. 
Finally, in Section~\ref{sec:proofoutline} we outline the proof. 


\section{\label{sec:model}Signal model and formal problem statement}

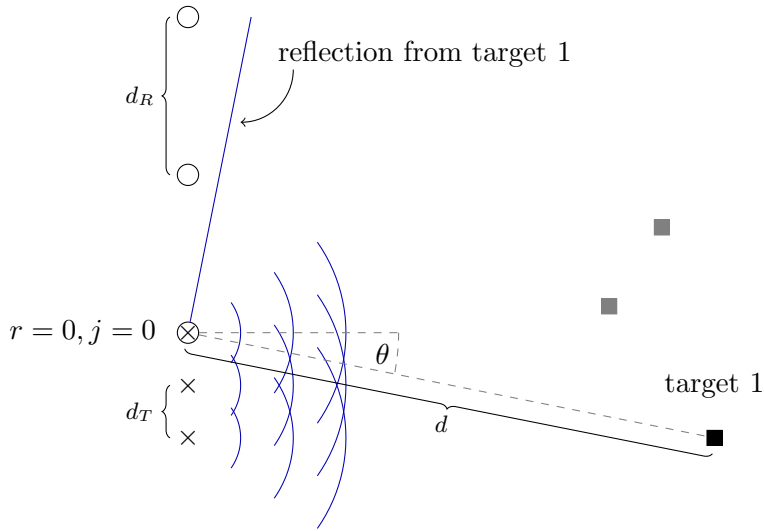
\begin{figure}[h!]
\begin{center}
\usetikzlibrary{intersections,decorations.pathreplacing}
\begin{tikzpicture}[scale=0.7]
\node (r1) at (0,0) [shape=circle,draw,inner sep =0.1cm] {};
\node (r2) at (0,3) [shape=circle,draw,inner sep =0.1cm] {};
\node at (0,6) [shape=circle,draw,inner sep =0.1cm] {};

\node at (0,0) [] {$\times$};
\node at (0,-1) [] {$\times$};
\node at (0,-2) [] {$\times$};

\foreach \y in {0,-1,-2} {
\draw [DarkBlue,domain=-35:35] plot ({1*cos(\x)}, {\y+sin(\x)});
\draw [DarkBlue,domain=-35:35] plot ({2*cos(\x)}, {\y+2*sin(\x)});
\draw [DarkBlue,domain=-35:35] plot ({3*cos(\x)}, {\y+3*sin(\x)});
};

\node (target) at (10,-2) [fill,shape = rectangle,draw,inner sep=0.1cm]{};

\node (target3) at (8,0.5) [gray, fill,shape = rectangle,draw,inner sep=0.1cm]{};
\node (target4) at (9,2) [gray,fill,shape = rectangle,draw,inner sep=0.1cm]{};

\path[color=gray,draw,dashed] (r1) -- (target);
\node at (10,-1) {target 1};

\path[color=gray,draw,dashed] (r1) -- (4,0);

\node at (3.7,-0.4) {$\theta$};
\draw[gray,draw,dashed] (4,0) arc(0:-12:4);

\draw [decorate,decoration={brace,amplitude=3pt},xshift=-4pt,yshift=0pt]
(-0.2,-2) -- (-0.2,-1) node [black,midway,xshift=-0.4cm] 
{\footnotesize $d_T$};

\draw [decorate,decoration={brace,amplitude=3pt},xshift=-4pt,yshift=0pt]
(-0.2,3) -- (-0.2,6) node [black,midway,xshift=-0.4cm] 
{\footnotesize $d_R$};

\draw [decorate,decoration={brace,amplitude=3pt},xshift=-4pt,yshift=0pt]
(10.1,-2.3) -- (0.08,-0.3) node [black,midway,yshift=-0.3cm,xshift=-0.1cm] 
{\footnotesize $d$};

\node (rft) at (4.5,5.3) {reflection from target 1};
\draw[->] (2,5) to [out=-90,in=0] (1,4);

\draw[DarkBlue] (r1) -- (1.2,6);

\node at (-2,0) {$r=0,j=0$};

\end{tikzpicture}
\end{center}
\caption[Principle of MIMO radar]{ \label{fig:geomrad}
Principle of MIMO radar: $\times$ and \tikz \node (r1) at (0,-0.05cm) [shape=circle,draw,inner sep =0.1cm] {}; correspond to transmit  and receive antennas. 
Throughout, we assume the spacing of the $N_T$ transmit and $N_R$ receive antennas to be $d_T= \frac{1}{2f_c}$ and $d_R= \frac{N_T}{2f_c}$, where $f_c$ is the carrier frequency. 
}
\end{figure}

We consider a MIMO radar with $N_T$ transmit and $N_R$ receive antennas that are colocated and lie in a plane along with $S$ targets, see Figure~\ref{fig:geomrad} for an illustration. 
While our results can be generalized to targets lying in three-dimensional space, we focus on targets lying in a plane, for simplicity. 
We assume that the targets are located in the far field of the array and let the transmit and receive antennas be uniformly spaced with spacings $\frac{1}{2 f_c}$ and $\frac{N_T}{ 2 f_c}$, respectively, where $f_c$ is the  carrier frequency. 
This spacing yields a uniformly spaced \emph{virtual array} with $N_T N_R$ antennas, and thus maximizes the number of virtual antennas achievable with $N_T$ transmit and $N_R$ receive  antennas \cite{friedlander_relationship_2009,strohmer_analysis_2014}. 
The (baseband) signal $y_r(t)$ at continuous time $t$ received by antenna $r=0,\hdots, N_R-1$, consists of the superposition of the reflections from the targets of the transmitted probing signals $x_j(t), j= 0,\hdots,N_T-1$, and is given by (see Appendix~\ref{sec:physrad} for more details), 
\begin{align}
y_r(t)
=
\sum_{k=0}^{\S-1}
b_k 
e^{i2\pi r N_T \beta_k}
\sum_{j=0}^{N_T-1} 
 e^{i2\pi j \beta_k }  x_j(t - \tauc_k) e^{i2\pi \nuc_k t}.
 \label{eq:iorelintro}
\end{align}
Here, $b_k \in \complexset$, $\beta_k \in [0,1]$, $\tauc_k$, and $\nuc_k$ are the attenuation factor, angle or azimuth parameter, delay, and Doppler shift associated with the $k$-th target. 
The parameters $\beta_k, \tauc_k,\nuc_k$ determine the angle ($\beta = -\sin(\theta)/2$ see Figure~\ref{fig:geomrad}), distance, and velocity of the $k$-th target relative to the radar. 
Locating the target therefore amounts to estimate the continuous parameters $b_k,\beta_k,\tauc_k,\nuc_k$ 
from the responses $y_r, r=0,\hdots,N_R-1$, to known and suitably selected probing signals $x_j$. 

In practice, however, the probing signals $x_j$ must be band-limited and approximately time-limited, and the responses $y_r$ can be observed only over a finite time interval. 
For concreteness, we assume that the $y_r$ are observed over an interval of length $T$ and that $x_j$ has bandwidth $B$ and is approximately supported on a time interval proportional to $T$. 
From the input-output relation~\eqref{eq:iorelintro}, it is evident that band- and approximate time-limitation of the $x_j$ implies that $y_r$ is band- and approximately time-limited as well---provided that the delay-Doppler pairs are compactly supported. 
This is indeed the case, due to path loss and finite velocity of the targets or objects in the scene \cite{strohmer_pseudodifferential_2006}. 
Formally, we assume that 
$
(\tauc_k, \nuc_k) \in [-T/2,T/2]\times[-B/2,B/2]
$. This is not restrictive since the region in the delay-Doppler plane where the delay-Doppler pairs are located can have area $BT \gg 1$, 
which is very large. 
In fact, a common assumption, not needed here, is that the delay-Doppler pairs lie in a region of area $\ll 1$ \cite{taubock_compressive_2010,bajwa_identification_2011}. 
Since the received signal $y_r$ is band-limited and approximately time-limited, by the $2WT$-Theorem \cite{slepian_bandwidth_1976,durisi_sensitivity_2012}, it is essentially characterized by on the order of $BT$-many coefficients. 
We therefore sample the received signals $y_r$ in the interval $[-T/2, T/2]$ at rate $1/B$, and collect the corresponding samples in the vectors\footnote{For simplicity we assume throughout that $\L = BT$ is an odd integer.} $\vy_r\in \complexset^{\L}$, $\L \defeq BT$, i.e., the 
$p$-th entry of $\vy_r$ is $[\vy_{r}]_p \defeq y_j(p/B)$, for $p = -\N,\hdots,\N, N \defeq \frac{L-1}{2}$. 
As detailed in \cite{heckel_super-resolution_2015}, those samples are given by\footnote{More precisely, equality~\eqref{eq:periorel} holds \emph{exactly} provided the signals $x_j$ are $T$-periodic on $\reals$, which, however, means that the signals $x_j$ are not time-limited. 
We hasten to add that if we let the signals $x_j$ be only partially periodic  
so that $x_j$ is essentially supported on an interval of length proportional to $T$, then equality~\eqref{eq:periorel} continues to hold \emph{approximately}. The corresponding relative error (for a random probing signal, as used here) is shown in \cite{heckel_super-resolution_2015} to decay as $1/\sqrt{\L}$ 
and is therefore negligible for large $\L$. 
}
\begin{align}
[\vy_r]_p
&= 
\sum_{k=0}^{\S-1}
b_k 
e^{i2\pi r N_T \beta_k } 
\sum_{j=0}^{N_T-1}
e^{i2\pi j \beta_k } 
[\mc F_{\nu_k}
\mc T_{\tau_k}
\vx_j ]_p,
\label{eq:periorel}
\end{align}
where $[\vx_j]_p \defeq x_j(p/B)$, 
$[\mc F_{\nu} \vx ]_p
\defeq [\vx]_p e^{i2\pi p \nu }$ and 
\begin{align}
[\mc T_{\tau} \vx ]_p
\!\defeq\!
\frac{1}{L}
\! \sum_{k=-\N}^{\N} \!
\!\left[\! 
\left(
\sum_{\ell=-\N}^{\N}
[\vx]_{\ell} e^{- i2\pi \frac{\ell k}{L}  }
\right)
\!e^{-i2\pi k  \tau }\! 
\right]
e^{i2\pi \frac{p k}{L}  }.
\label{eq:deftimefreqshifts}
\end{align}
Here, we defined 
 the time-shifts $\tau_k \defeq  \tauc_k/T$ and frequency-shifts $\nu_k \defeq \nuc_k/B$. Since $(\tauc_k, \nuc_k) \in [-T/2,\allowbreak T/2] \allowbreak \times[-B/2,B/2]$ we have $(\tau_k, \nu_k) \in [-1/2,1/2]^2$. Since $\mc T_{\tau}\vx$ and $\mc F_{\nu}\vx$ are $1$-periodic in $\tau$ and $\nu$, we can assume in the remainder of the paper that $(\tau_k, \nu_k) \in [0,1]^2$. 
The operators $\mc T_{\tau}$ and $\mc F_{\nu}$ can be interpreted as fractional time and frequency shift operators in $\complexset^\L$. If the $(\tau_k,\nu_k)$ lie on a $(1/L,1/L)$ grid, $\mc F_{\nu}$ and $\mc T_{\tau}$ reduce to the ``natural'' time frequency shift operators in $\complexset^\L$, i.e., $[\mc T_{\tau} \vx ]_p = x_{p - \tau \L}$ and $[\mc F_{\nu} \vx ]_p = x_p e^{i2\pi p \frac{\nu\L}{\L} }$. 
The definition of a time shift in \eqref{eq:deftimefreqshifts} as taking the Fourier transform, 
modulating the frequency, and taking the inverse Fourier transform is a very natural definition of a \emph{continuous} time-shift $\tau_k \in [0,1]$ of a \emph{discrete} vector $\vx = \transp{[x_{-\N},\hdots,x_{\N}]}$. 

We have reduced the problem of identifying the locations of the targets 
under the constraints that the probing signals $x_j$ are band-limited and the responses $y_r$ are observed over a finite time interval only, 
to the estimation of the parameters $b_k \in \complexset$, $(\beta_k,\tau_k, \nu_k )\in [0,1]^3, k=0,\hdots,\S-1$ from the samples $[\vy_r]_p, r = 0,\hdots,N_R-1, p=-\N,\hdots,\N$, in the input-output relation~\eqref{eq:periorel}. 
We call this the super-resolution MIMO radar problem. 

\section{\label{sec:atomicnmin}Recovery via atomic norm minimization}

We next formally present our recovery algorithm. 
To this end, we first define for convenience the vector $\vr \defeq [\beta,\tau,\nu]$, and write the input-output relation \eqref{eq:periorel} in matrix-vector form:
\begin{align}
\vy = \mAA \vz, 
\quad \vz = \sum_{k=0}^{\S-1} b_k \atom(\vr_k).
\label{eq:syseqinw}
\end{align}
Here, $\vy \defeq \transp{[ \transp{\vy}_0,\hdots, \transp{\vy}_{N_R-1} ] }$, 
where $\atom(\vr) \in \complexset^{\L^2 N_TN_R}$ has entries\footnote{Here, and in the following we use for convenience a three dimensional index to refer to entries of the vector $\atom$.} $[\atom(\vr)]_{(v,k,p)} = e^{i2\pi(v \beta + k\tau + p\nu)}, v=0,\hdots,N_T N_R-1$, $k,p=-\N,\hdots,\N$, and  $\mAA \in \complexset^{N_R \L \times N_R N_T \L^2}$ is defined as follows. The expression $w_{r,p} \defeq e^{i2\pi r N_T \beta } 
\sum_{j=0}^{N_T-1}
e^{i2\pi j \beta } 
[\mc F_{\nu}
\mc T_{\tau}
\vx_j ]_p$ in \eqref{eq:periorel} can be written as
\[
w_{r,p}
=
\sum_{j=0}^{N_T-1}
\sum_{k = -\N}^{\N} 
a_{p,k,j}
   e^{i2\pi (k \tau + p \nu + (j+N_T r) \beta)},
\]
with $a_{p,k,j} = \frac{1}{\L} \sum_{\ell = -\N}^{\N}
  [\vx_j]_\ell
    e^{i2\pi (\ell-p) \frac{k}{L} }$. 
Let $\atom_{p,j} \in \complexset^\L$ be the vector with $k$th entry $[\atom_{p,j}]_k = a_{p,k,j}$, $k=-\N,\hdots,\N$, and let $\mA_j \in \complexset^{\L \times \L^2}$ be the block-diagonal matrix with $\transp{\atom}_{p,j}$ on its $p$th diagonal, $p=-\N,\hdots,\N$. With this notation, $\mAA$ is defined as the block-diagonal matrix with the matrix $[\mA_0,\hdots,\mA_{N_T-1}] \in \complexset^{\L \times N_T \L^2}$ on its diagonal, for all $N_R$ blocks on the diagonal. With this notation, \eqref{eq:periorel} becomes \eqref{eq:syseqinw}. 

The significance of the representation \eqref{eq:syseqinw} is that recovery of the $b_k,\vr_k = [\beta_k,\tau_k,\nu_k]$ from $\vz$ is a 3D line spectral estimation problem that can be solved with standard spectral estimation techniques such as Prony's method \cite{stoica_spectral_2005}. 
Therefore, we only need to recover $\vz \in \complexset^{N_RN_T \L^2}$ from $\vy \in \complexset^{N_R\L}$. 
To do so, we use that $\vz$ is a sparse linear combination of  atoms in the set $\setA \defeq \{ \atom(\vr), \vr \in [0,1]^3\}$. 
A regularizer that promotes such a sparse linear combination is the atomic norm induced by these signals~\cite{chandrasekaran_convex_2012}, defined as 
$
\norm[\setA]{\vz} 
\defeq \inf_{b_k \in \complexset, \vr_k \in [0,1]^3} \left\{ \sum_k |b_k| \colon \vz = \sum_k b_k \atom(\vr_k) \right\}. 
$
We estimate $\vz$ by solving the basis pursuit type atomic norm minimization problem problem  
\newcommand{\AN}{\mathrm{AN}}
\begin{align}
\AN(\vy) \colon \;\; \underset{\minlet{\vz}  }{\text{minimize}} \,  \norm[\setA]{\minlet{\vz} } \; \text{ subject to } \; \vy = \mAA \minlet{\vz}.
\label{eq:primal}
\end{align}

To summarize, we estimate the $b_k,\vr_k$ from $\vy$ by i) solving $\AN(\vy)$ in order to obtain $\vz$, ii) estimating the $\vr_k$ from $\vz$ by solving the corresponding 3D-line spectral estimation problem, and iii) solving the linear system of equations 
$
\vy = \sum_{k=0}^{\S-1} b_k \mAA \atom(\vr_k)
$ for the $b_k$. 

We remark that the $\vr_k$ may be obtained more directly from a solution to the dual of \eqref{eq:primal}; 
see \cite[Sec.~3.1]{bhaskar_atomic_2012}, \cite[Sec.~4]{candes_towards_2014}, \cite[Sec.~2.2]{tang_compressed_2013}, and \cite[Sec.~6]{heckel_super-resolution_2015} for details on this approach applied to related problems. 

Since computation of the atomic norm involves taking the infimum over infinitely many parameters, finding a solution to $\AN(\vy)$ may appear to be daunting. 
For the 1D case (i.e., only angle, time, or frequency shifts), the atomic norm can be characterized in terms of linear matrix inequalities (LMIs) \cite[Prop.~2.1]{tang_compressed_2013}. This characterization is based on the Vandermonde decomposition lemma for Toeplitz matrices,  
and allows to formulate the atomic norm minimization program as a semidefinite program  that can be solved in polynomial time. 
While this lemma generalizes to higher dimensions \cite[Thm.~1]{yang_vandermonde_2015}, it fundamentally comes with a rank constraint 
that appears to prohibit an straightforward characterization of the atomic norm in terms of LMIs. 
Nevertheless, based on \cite[Thm.~1]{yang_vandermonde_2015}, one can obtain a semidefinite programming (SDP) \emph{relaxation} of $\AN(\vy)$, which can be solved in polynomial time. 
Similarly, a solution of the dual of $\AN(\vy)$ can be found with a SDP relaxation. 
Since the computational complexity of the corresponding semidefinite programs is quite large, we will not dive into the details of those SDP relaxations. 
Instead, we show in Section \ref{sec:discretesuperes} that the $\vr_k$ can be recovered on an arbitrarily fine grid via $\ell_1$-minimization. 
While this leads to a gridding error, the grid may be chosen sufficiently fine for the gridding error to be negligible compared to the error induced by additive noise (in practice, there is typically additive noise). 


\section{\label{sec:mainares}Main results for atomic norm minimization}

Throughout, we take the probing signals to be random by choosing its samples, i.e., the entries of the $\vx_j$ as i.i.d.~Gaussian (or sub-Gaussian) zero-mean random variables with variance $1/(N_T \L)$. 
Our main results is stated next. 

\begin{theorem} 
Assume $\L = 2\N+1 \geq 1024$, $N_T N_R \geq 1024$, and suppose we observe 
$
\vy = \mAA \vz, 
\quad \vz = \sum_{k=0}^{\S-1} b_k \atom(\vr_k)
$
where $\sign(b_k)$ 
is chosen independently from symmetric distributions on the complex unit circle and the $\vr_k = [\beta_k,\tau_k,\nu_k]$ are arbitrary triplets obeying the minimum separation condition  
\begin{align}
|\beta_k - \beta_{k'}| \geq \frac{10}{N_T N_R- 1} \quad \text{or}\quad
 |\tau_k - \tau_{k'}| \geq \frac{5}{\N} 
\quad\text{or}\quad
|\nu_k - \nu_{k'}|  \geq \frac{5}{\N}
, \quad \text{for all $k,k'\colon k\neq k'$}. 
\label{eq:minsepcond}
\end{align}
Here, $|\beta_k - \beta_{k'}|$ is the wrap-around distance on the unit circle. For example, $|3/4-1/2|=1/4$ but $|5/6-1/6|=1/3\neq 2/3$. 
Choose $\delta>0$ and assume that
\begin{align}
S \le c \min(L, N_T N_R) / \log^3\left(L/\delta \right)
\label{eq:sleqb}
\end{align}
where $c$ is a numerical constant. Then, with probability at least $1-\delta$, $\vz$ is the unique minimizer of $\AN(\vy)$ in \eqref{eq:primal}. 
\label{thm:mainres}
\end{theorem}


Theorem~\ref{thm:mainres} guarantees that, with high probability, the parameters $b_k,\vr_k$ can be recovered perfectly from the observation $\vy$ by solving a convex program (recall that the parameters $b_k, \vr_k$ can be obtained from $\vz$), provided that the locations $\vr_k = [\beta_k,\tau_k,\nu_k]$ are sufficiently separated in \emph{either} angle, time, or frequency, and provided that the total number of targets satisfies condition \eqref{eq:sleqb}. 
Note that, translated to the physical parameters $\tauc_k,\nuc_k$, the minimum separation condition \eqref{eq:minsepcond} becomes: 
\[
|\beta_k - \beta_{k'}| \geq \frac{10}{N_T N_R- 1}
\quad \text{or} \quad 
 |\tauc_k - \tauc_{k'}| \geq \frac{10.01}{B}
\quad \text{or} \quad 
|\nuc_k - \nuc_{k'}|  \geq \frac{10.01}{T}, \quad \text{for all $k,k'\colon k\neq k'$}.
\]

Theorem~\ref{thm:mainres} is essentially optimal in the number of targets that can be located, since $\S$ can be linear---up to a log-factor---in $\min(\L, N_T N_R)$, and $\S \leq \min(\L, N_T N_R)$ is a necessary condition to uniquely recover the attenuation factors $b_k$ even if the locations $\vr_k$ are known. 
To see this, note that for the linear system of equations  \eqref{eq:syseqinw} to have a unique solution, 
the vectors $\mAA \atom(\vr_k)$ must be linearly independent. 
If $\beta_k=0$, for all $k$, or if $\tau_k=0$ and $\nu_k=0$, for all $k$, the vectors $\mAA \atom(\vr_k), \vr_k = (\beta_k, \tau_k, \nu_k), k=0,\hdots,\S-1$ 
can only be linearly independent provided that $\S \leq \L$ and $\S \leq N_T N_R$, respectively. 
This is seen from 
\[
\mAA \atom(\vr)
=
\begin{bmatrix}
e^{i2\pi 0 \beta }  \sum_{j=0}^{N_T-1} e^{i2\pi j \beta} \mc F_{\nu}
\mc T_{\tau} \vx_j  \\
\vdots  \\
e^{i2\pi N_T (N_R-1) \beta }  \sum_{j=0}^{N_T-1} e^{i2\pi j \beta} \mc F_{\nu}
\mc T_{\tau} \vx_j 
\end{bmatrix}.
\]

Regarding the minimum separation condition, we note that some sort of separation between the $(\beta_k,\tau_k,\nu_k)$ is necessary for \emph{stable} recovery. 
This follows from the simpler problem of line spectral estimation, that is obtained from our setup by setting $\beta_k =0, \tau_k=0$ for all $k$, being 
ill posed if the $\nu_k$ are clustered closely together. 
Specifically, suppose $\S'$ frequencies $\nu_k$ are in an interval of length smaller than $\frac{2\S'}{\L}$. 
For $\S'$ large, the problem of recovering the $(b_k,\nu_k)$ is extremely ill-posed \cite[Thm.~1.1]{donoho1992superresolution}, \cite{morgenshtern2014stable}, \cite[Sec.~1.7]{candes_towards_2014}. 
Condition~\eqref{eq:minsepcond} 
allows us to have  $0.2\,\S'$ time-frequency shifts in an interval of length $\frac{2\S'}{\L}$, which is optimal up to the constant $0.2$. 
%
%

Recall that the complex-valued coefficients $b_k$ in the radar model \eqref{eq:iorelintro} describe the attenuation factors. 
A standard modeling assumption in wireless communication and radar~\cite{bello_characterization_1963} is that the $b_k$ are complex Gaussian distributed. 
Under this model, the random sign assumption in Theorem~\ref{thm:mainres} is satisfied. 
However, we believe that the random sign assumption is not necessary for our result to hold.

Theorem~\ref{thm:mainres} is proven by constructing an appropriate dual certificate; the existence of this  certificate guarantees that the solution to $\AN(\vy)$ in \eqref{eq:primal} is $\vz$, as formalized by 
Proposition~\ref{prop:dualmin} below. 
Proposition~\ref{prop:dualmin} is a consequence of strong duality, and well known for the discrete setting from the compressed sensing literature \cite{candes_robust_2006}. 
The proof is standard, see e.g., \cite[Proof of Prop.~2.4]{tang_compressed_2013}. 

\begin{proposition} 
Let $\vy=\mAA \vz$ with $\vz =   \sum_{k=0}^{\S-1}  b_k  \atom(\vr_k)$. If there exists a dual certificate
$
\tilde Q(\vr) =  \innerprod{\vq}{ \mAA \atom(\vr) }
$ 
with complex coefficients $\vq \in \complexset^{N_R \L}$ such that 
\begin{align}
\tilde Q(\vr_k) = \sign(b_k), \text{ for all $k$, and } |\tilde Q(\vr)| < 1 \text{ for all } \vr \in [0,1]^3 \setminus \{\vr_0,\hdots,\vr_{\S-1}\}, 
\label{eq:dualpolyinatmincon}
\end{align}
then $\vz$ is the unique minimizer of $\AN(\vy)$. 
\label{prop:dualmin}
\end{proposition}

\section{\label{sec:discretesuperes} Recovery on a fine grid}

An practical approach to estimate the parameters $\vr_k$ from the received signals $\vy_r$ in the input-output relation~\eqref{eq:periorel}, promoted in \cite{tang_sparse_2013}, is to suppose the angle-time-frequency triplets lie on a \emph{fine} grid, and solve the problem on that grid. 
In general this leads to a gridding error, that, however, becomes small as the grid gets finer. 
We next discuss the corresponding (discrete) sparse signal recovery problem. 

Suppose the $(\beta_k,\tau_k,\nu_k)$ lie on a grid with spacing $(1/K_1,1/K_2,1/K_3)$, where $K_1,K_2,K_3$ are integers obeying $K_1 \geq N_TN_R$, $K_2,K_3 \geq \L = 2 \N +1$. 
With this assumption, the super-resolution MIMO radar problem reduces to the recovery of the sparse (discrete) signal $\vs\in \complexset^{K_1 K_2 K_3}$ 
from the measurement
$
\vy = \mR \vs
$
where $\mR \in \complexset^{N_R \L  \times K_1 K_2 K_3}$ is the matrix with $(n_1,n_2,n_3)$-th column given by $\mAA \atom(\vr_\vn)$, $\vr_\vn = (n_1/\K_1, n_2/\K_2, n_3/\K_3)$. 
Note that the non-zeros of $\vs$ and its indices correspond to the attenuation factors $b_k$ and the locations $\vr_k$ on the grid. 
A standard approach to the recovery of the sparse signal $\vs$ from the underdetermined linear system of equations $\vy = \mR \vs$ is to solve the following convex program:
\begin{align}
	\mathrm{L1}(\vy) \colon \;\;
\underset{\minlet{\vs}}{\text{minimize}} \; \norm[1]{\minlet{\vs}} \text{ subject to } \vy = \mR \minlet{\vs}.
\label{eq:l1minmG}
\end{align}
Below is our main result for recovery on the fine grid.   

\begin{theorem}  
Assume $\L = 2\N+1 \geq 1024$, $N_T N_R \geq 1024$, and suppose we observe $\vy = \mR \vs$, 
where $\vs$ is a sparse vector with non-zeros indexed by the support set $\mc S \subseteq [K_1]\times [K_2] \times [K_3]$. 
Suppose that those indices satisfy the minimum separation condition: For all triplets $(n_1,n_2,n_3), (n_1',n_2',n_3') \in \mc S$, 
\[
\frac{|n_{1} - n_{1}'|}{K_1} \geq \frac{10}{N_T N_R- 1} 
\quad \text{or}\quad
\frac{|n_{2} - n_{2}'|}{K_2} \geq \frac{5}{\N}
\quad\text{or}\quad
\frac{|n_{3} - n_{3}'|}{K_3}  \geq \frac{5}{\N}.
\]
Moreover, we assume that the signs of the non-zeros of $\vs$ are chosen independently from symmetric distributions on the complex unit circle. 
Choose $\delta>0$ and assume
$
S \le c \min(L, N_T N_R) / \log^3\left(\L/\delta \right) 
$
where $c$ is a numerical constant. Then, with probability at least $1-\delta$, $\mathbf{s}$ is the unique minimizer of $\mathrm{L1}(\vy)$ in \eqref{eq:l1minmG}. 
\label{cor:discretesuperres}
\end{theorem}

Note that Theorem~\ref{cor:discretesuperres} does not impose any restriction on $K_1,K_2,K_3$, in particular they can be arbitrarily large.  
The proof of Theorem~\ref{cor:discretesuperres} is closely linked to that of Theorem~\ref{thm:mainres}. 
Specifically, 
the existence of a certain dual certificate guarantees that $\vs$ is the unique minimizer of $\mathrm{L1}(\vy)$. 
The dual certificate is obtained directly from the dual certificate for the continuous case (i.e., from $\tilde Q(\vr)$ in Proposition~\ref{prop:dualmin}). 


\section{\label{sec:numres} Numerical results and robustness}

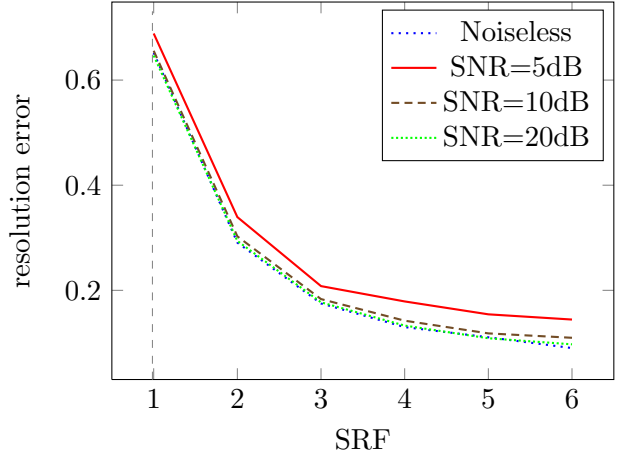
\begin{figure}
\begin{center}
 \begin{tikzpicture}[scale=1,thick] 
 \begin{axis}[xlabel={$\text{SRF}$}, ylabel = {resolution error},  xtick={1,2,3,4,5,6},
 width = 0.5\textwidth, height=0.4\textwidth]   
 \addplot +[thick,mark=none, dotted] table[x index=0,y index=1]{./dat/offgrid.dat}; 
 \addlegendentry{Noiseless}
 \addplot +[thick,mark=none] table[x index=0,y index=2]{./dat/offgrid.dat};
 \addlegendentry{SNR=5dB}
  \addplot +[thick,mark=none,densely dashed] table[x index=0,y index=3]{./dat/offgrid.dat}; 
 \addlegendentry{SNR=10dB}
  \addplot +[thick,mark=none, green, densely dotted] table[x index=0,y index=4]{./dat/offgrid.dat};
 \addlegendentry{SNR=20dB} 
  \end{axis}
  \draw[very thin,dashed,gray] (0.54,0) -- (0.54,4.9);
 \end{tikzpicture}
 \end{center}
 \caption{\label{fig:realrecov} 
 Resolution error for the recovery of $\S = 5$ targets from the samples $\mathbf y$ with and without additive Gaussian noise $\mathbf n$ of a certain signal-to-noise ratio $\text{SNR} = \| \mathbf y \|^2_2/ \| \mathbf n \|^2_2$, for varying super-resolution factors (SRFs).  
 The resolution error is defined as the average over $( N_T^2 N_R^2(\hat \beta_k - \beta_k)^2 + \L^2(\hat \tau_k - \tau_k)^2 + \L^2(\hat \nu_k - \nu_k)^2)^{1/2}$, $k=0,\ldots,\S-1$,  where $(\hat \beta_k,\hat \tau_k, \hat \nu_k)$ are the locations obtained by solving \text{L1-ERR}.   
}
\end{figure}

We briefly evaluate numerically the resolution obtained by our approach, and demonstrate that it is robust to noise.
We set $N_T = 3, N_R = 3$, and $\L = 41$, and draw $S=5$ target locations $(\beta_k,\tau_k,\nu_k)$ uniformly at random from $[0,1] \times [0,2/\sqrt{\L}]^2$. Moreover, we choose $K_1 = \SRF N_T N_R, K_2 = \SRF \L$, and $K_3 = \SRF \L$, where $\SRF\geq 1$ can be interpreted as a super-resolution factor as it determines by how much the $(1/K_1,1/K_2,1/K_3)$ grid is finer than the original, coarse grid $(1/(N_TN_R),1/\L,1/\L)$. 
 To account for additive noise, we solve the following modification of $\mathrm{L1}(\vy)$ in \eqref{eq:l1minmG}
\begin{align}
\text{L1-ERR}\colon \underset{\minlet{\vs}}{\text{minimize}} \; \norm[1]{\minlet{\vs}} \text{ subject to } 
\norm[2]{\vy - \mR \minlet{\vs} }^2 \leq \delta,
\label{eq:BDDN}
\end{align}
with $\delta$ chosen on the order of the noise variance. 
There are two error sources incurred by this approach: the gridding error obtained by assuming the points lie on a grid with spacing $(1/K_1,1/K_2,\allowbreak1/K_3)$---which decreases in $\SRF$---and the additive noise error. 
The results, depicted in Figure~\ref{fig:realrecov}, show that the target resolution of the super-resolution approach is significantly better than that of the compressed sensing based approach \cite{dorsch_refined_2015,strohmer_adventures_2015} corresponding to recovery on the coarse grid i.e., $\SRF=1$. 
Moreover, the results show that our approach is robust to noise. 
 

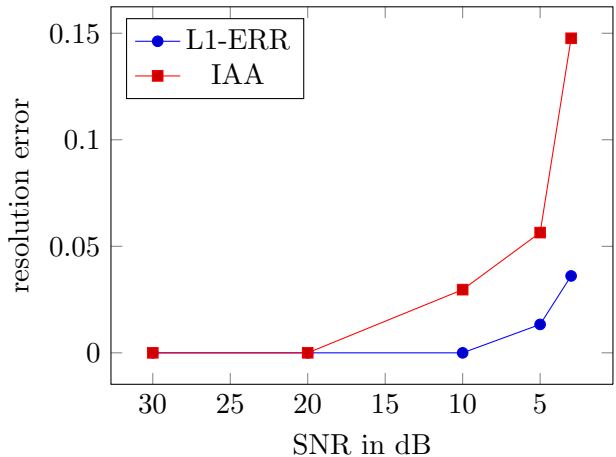
\begin{figure}[!ht]
\centering
\begin{tikzpicture}    
\begin{axis}[
xlabel = SNR in dB,
ylabel=resolution error,
width = 0.5\textwidth,
height=0.4\textwidth,
x dir=reverse,
legend pos=north west,
yticklabel style={/pgf/number format/fixed,
                  /pgf/number format/precision=3}
]

\addplot +[blue] table[x index=0,y index=1]{./dat/resultsSRF_nfp.dat};
\addlegendentry{ L1-ERR };

\addplot +[red] table[x index=0,y index=2]{./dat/resultsSRF_nfp.dat};
\addlegendentry{IAA};

\end{axis}
\end{tikzpicture}

\caption{
\label{fig:cmpmusicconvex}
Resolution error (smaller is better) of L1-ERR and IAA applied to $\vy + \vn$, where $\vn \in \complexset^{N_R\L}$ is additive Gaussian noise, such that the signal-to-noise ratio is 
 $\text{SNR} \defeq \norm[2]{\vy}^2 / \norm[2]{ \vn }^2$. 
 As before, the resolution error is defined as $( N_T^2 N_R^2(\hat \beta_k - \beta_k)^2 + \L^2(\hat \tau_k - \tau_k)^2 + \L^2(\hat \nu_k - \nu_k)^2)^{1/2}$, 
where $(\hat \beta_k,\hat \tau_k, \hat \nu_k)$ are the locations obtained by solving \text{L1-ERR}.  
}

\end{figure}

We next compare our approach to the  Iterative Adaptive Approach (IAA)~\cite{yardibi_source_2010}, proposed for MIMO radar in the paper~\cite{roberts_iterative_2010}. IAA is based on weighted least squares and has been proposed in the array processing literature. IAA  can work well even with one snapshot only and can therefore be directly applied to the MIMO super-resolution problem. 
However, to the best of our knowledge, no analytical performance guarantees are available in the literature that attest IAA similar performance than the $\ell_1$-minimization based approach. 
We compare the IAA algorithm \cite[Table II, ``The IAA-APES Algorithm'']{yardibi_source_2010} to L1-ERR, for a problem with parameters 
$N_T = 3, N_R = 3$, and $\L = 41$, as before, but with $\SRF = 3$ and $(\beta_k,\tau_k,\nu_k) = (k/(N_R N_t), k/\L, k/\L), k=0,\ldots,\S-1$, so that the 
location parameters lie on the fine grid, and are separated. 
As before, we draw the corresponding attenuation factors $b_k$ i.i.d.~uniformly at random from the complex unit disc. 
Our results, depicted in Figure \ref{fig:cmpmusicconvex}, show that L1-ERR performs better in this experiment than IAA, in particular for small signal-to-noise ratios.

\section{\label{sec:proofoutline}Proof outline}

Theorem~\ref{prop:dualmin} and Theorem~\ref{cor:discretesuperres} both follow from the existence of a dual certificate as specified in Proposition~\ref{prop:dualmin}. 
Our construction of the dual certificate $\tilde Q$ 
 is inspired by the construction of related certificates in \cite{candes_towards_2014,tang_compressed_2013,heckel_super-resolution_2015}.
From 
$
\tilde Q(\vr) 
= \innerprod{\vq}{ \mAA \atom(\vr)}
= \innerprod{\herm{\mAA}\vq}{\atom(\vr)},
$ 
it is seen that $\tilde Q$ is a $3D$-trigonometric polynomial in the variables $\beta,\tau,\nu$ with coefficient vector $\herm{\mA} \vq$ 
(recall that the entries of $\atom(\vr), \vr = (\beta,\tau,\nu)$ are given by $[\atom(\vr)]_{(v,k,p)} = e^{i2\pi(v \beta + k\tau + p\nu)}$). 
To build $\tilde Q$ we therefore need to 
construct a 3D-trigonometric polynomial that satisfies condition~\eqref{eq:dualpolyinatmincon}, and whose coefficients are constraint to be of the form $\herm{\mA} \vq$. 
Since the $\vx_j$ are random, and $\mA$ depends on the $\vx_j$, $\tilde Q$ is a \emph{random} trigonometric polynomial. 

For notational simplicity, we assume that $N_R N_T = \L$, and define the polynomial 
$Q(\vr) 
= e^{i2\pi N\beta} \tilde Q(\vr)
\big<\herm{\mA}\vq , \atomcen(\vr) \big>
$, 
where the entries of the vector $\atomcen(\vr)$ are given by $[\atomcen(\vr)]_{(v,k,p)} = e^{i2\pi(v \beta + k\tau + p\nu)}, \allowbreak v,k,p=-\N,\hdots,\N$. 
With this notation, condition \eqref{eq:dualpolyinatmincon} on $\tilde Q$ is equivalent to $Q$ obeying
\begin{align}
&Q(\vr) = u_k, \quad u_k \defeq e^{i2\pi N \beta_k} \sign(b_k), \text{ for all $k$}  \nonumber \\ 
&\text{and }|Q(\vr)| < 1, \text{ for all } \vr \in [0,1]^3 \setminus \{\vr_0,\hdots, \vr_{\S-1}\}. 
\label{eq:intboundcondpro}
\end{align}
We construct $Q$ explicitly. 
It is instructive to first consider the construction of a \emph{deterministic} 3D trigonometric polynomial $\bar Q(\vr) = \big<\bar \vq,\atomcen(\vr)\big>$ with unconstraint, deterministic coefficients $\bar \vq$, 
that satisfies condition~\eqref{eq:intboundcondpro}, but whose coefficients $\bar \vq \in \complexset^{L^3}$ are \emph{not} constraint to be of the form $\herm{\mA} \vq$. 
Such a construction has been established (provided a minimum separation condition on the $\vr_k$ holds) by Cand\`es and Fernandez-Granda \cite[Prop.~2.1, Prop.~C.1]{candes_towards_2014} for the 1D and 2D case; the 3D case is treated analogously. 
To construct $Q$, Cand\`es and Fernandez-Granda \cite{candes_towards_2014} interpolate the points $u_k$ with a fast-decaying kernel $\bar G(\vr) \defeq F(\beta) F(\tau) F(\nu)$ and slightly adopt this interpolation near the $\vr_k$ with the partial derivatives $\bar G^{(n_1,n_2,n_3)}(\vr) \defeq 
\frac{\derd^{n_1} }{ \derd \beta^{n_1}}
\frac{\derd^{n_2} }{ \derd \tau^{n_2}} \frac{\derd^{n_3} }{ \derd \nu^{n_3}}  
\bar G(\vr)$ to ensure that local maxima are achieved at the $\vr_k$: 
\begin{align}
\bar Q(\vr) 
= \sum_{k=1}^\S & \bar \alpha_k \bar G(\vr-\vr_k) + \bar \alpha_{1k} \bar G^{(1,0,0)}(\vr - \vr_k) 
+   \bar\alpha_{2k} \bar G^{(0,1,0)}(\vr - \vr_k)
+  \bar\alpha_{3k} \bar G^{(0,0,1)}(\vr - \vr_k). 
\label{eq:detintpolC}
\end{align}
Here, $\FK$ is the squared Fej\'er kernel which is a certain trigonometric polynomial with coefficients $g_k$, i.e.,  
$
 \FK(t) =  \sum_{k=-\N}^{\N}  g_k e^{i2\pi t k}.
$ 
Shifted versions of $F$ (i.e., $F(t - t_0)$) and the derivatives of $\FK$ are also 1D trigonometric polynomials of degree $\N$, therefore $\bar G$, its partial derivatives, and shifted versions thereof are 3D trigonometric polynomials of the form $\big<\bar \vq, \atomcen(\vr)\big>$. 
The construction of $\bar Q$ is concluded by showing that the coefficients $\bar \alpha_k,\bar \alpha_{1k},\bar \alpha_{2k},\bar \alpha_{3k}$, can be chosen such that $\bar Q$ reaches global maxima at the $\vr_k$. 

Our construction of $Q$ follows a similar program.  
Specifically, we interpolate the points $u_k$ at $\vr_k$ with the functions 
$
G_{\vn}(\vr,\vr_k) = 
\big<
\mA \vg_{\vn}(\vr_k) ,
\mA \atomcen(\vr) 
\big>. 
$
Here, $\vg_{\vn}(\vr_k), \vn = (n_1,n_2,n_3)$ is the vector with $(v,k,p)$-th coefficient given by
$
g_v g_k g_p
(i2\pi v)^{n_1}(i2\pi k)^{n_2} (i2\pi p)^{n_3}
e^{-i2\pi (\beta v + \tau k + \nu p)}, 
$
where the $g_k$ are the coefficients of the squared Fej\'er kernel $F$. 
With this definition, we have 
$
\EX{G_{\vn}(\vr, \vr_k)} 
=  \bar G^{\vn}(\vr-\vr_k). 
$ 
This follows from $\EX{\herm{\mA} \mA } =\mI$, not shown here. 
Moreover, $G_{\vn}(\vr, \vr_k)$ concentrates around $\bar G^{\vn}(\vr-\vr_k)$. 
We construct $Q$ by interpolating the $u_k$ at $\vr_k$ with $G_{(0,0,0)}(\vr,\vr_k), k=0,\hdots,\S-1$, 
and slightly adopt this interpolation near the $\vr_k$ with linear combinations of $G_{(1,0,0)}(\vr,\vr_k), G_{(0,1,0)}(\vr,\vr_k)$, and $G_{(0,0,1)}(\vr,\vr_k)$, in order to ensure that local maxima of $Q$ are achieved exactly at the $\vr_k$. 
Specifically, we set 
\begin{align}
Q(\vr) = \sum_{k=1}^\S 
& \alpha_k G_{(0,0,0)}(\vr,\vr_k)
+ \alpha_{1k} G_{(1,0,0)}(\vr,\vr_k) 
+ \alpha_{2k} G_{(0,1,0)}(\vr,\vr_k)
+ \alpha_{3k} G_{(0,0,1)}(\vr,\vr_k).
\label{eq:dualpolyorig}
\end{align}  
Note that $Q(\vr)$ is a linear combination of the functions $G_{\vm}(\vr,\vr_k)$, and by definition of the $G_{\vm}(\vr,\vr_k)$ it obeys $\innerprod{\herm{\mA}\vq}{\atomcen(\vr)}$, for some $\vq$, as desired. 
The proof is concluded by showing that, with high probability, there exists a choice of coefficients $\alpha_k, \alpha_{1k}, \alpha_{2k}$ and $\alpha_{3k}$ such that $Q$ reaches global maxima at the $\vr_k$ and $Q(\vr_k) = u_k$, for all $k$. 
For this argument to work, the particular choice of $G_{\vm}(\vr, \vr_k)$ is crucial; the main ingredients are that $G_{\vm}(\vr, \vr_k)$ concentrates around $\bar G(\vr - \vr_k)$, and certain properties of $\bar G$ and $\bar Q$. 


\section*{Funding and Acknowledgements}
RH was supported by the Swiss National Science Foundation under grant P2EZP2\_159065, and would like to thank Mahdi Soltanolkotabi and Veniamin Morgenshtern 
for helpful discussions.

\printbibliography


\appendices

\section{Signal Model}
\label{sec:physrad}


In this appendix, we provide additional details on the input-output relation \eqref{eq:iorelintro}. 
Recall that we consider a MIMO radar with $N_T$ transmit antennas and $N_R$ receive antennas that are located on a line in a two-dimensional plane together with the targets, as displayed in Figure~\ref{fig:geomrad}. 
We assume that the targets are located in the far field of the array, i.e., the distance of the target to the antenna array is much larger than the largest distance between two antennas. 
As a consequence, propagating waves appear planar and the angle between the target and each antenna is (approximately) the same. 
For concreteness, we consider a setup where the transmit and receive antennas are uniformly spaced with spacings $d_T$ and $d_R$, respectively. 

Consider a single target first. 
The $j$-th antenna transmits the signal $x_j(t) e^{i2\pi f_c t}$, where $f_c$ is the carrier frequency. 
This signal propagates to the target, which we assume to be a point scatterer, gets reflected, and propagates back to the $r$-th receiver. 
From Figure~\ref{fig:geomrad}, we see that the corresponding delay is, as a function of the angle between antennas and the target, $\theta$,  distance to the target, $d$, and the speed of light, $c$, given by
\[
\tilde \tau 
\defeq
\frac{ 2d }{c}
+
\frac{\sin(\theta)(d_T j + d_R r )}{c}
= 
 \tauc - \beta  \frac{2(d_T j + d_R r ) }{c}. 
\]
For the second equality, we defined the angle parameter $\beta \defeq -\sin(\theta)/2$ and the delay $\tauc \defeq \frac{2d}{c}$. 
Taking the Doppler shift into account, the reflection of the $j$-th probing signal 
received by the $r$-th receive antenna is given by 
\begin{align}
\tilde b x_j(t   - \tilde \tau ) 
e^{i2\pi (f_c + \nuc ) \left( t -  \tilde \tau \right)}.
\label{eq:exresp}
\end{align}
Here, $\tilde b\in \complexset$ is the attenuation factor associated with the target, and $\nuc \defeq \frac{2 v }{c} f_c$ is the Doppler shift, which is a function of the relative velocity, $v$, of the object. 
By choosing the antenna spacing as $d_T = \frac{1}{2f_c}$ and $d_R =\frac{N_T}{2f_c}$, 
the reflection of the $j$-th probing signal received by the $r$-th receive antenna in \eqref{eq:exresp} becomes 
\[
\tilde b x_j( t - \tilde \tau ) 
e^{i2\pi (f_c + \nuc) \left( t -  \tauc \right)}
e^{i2\pi (f_c + \nuc) 
\beta \frac{j + r N_T}{f_c}       
}
\approx 
\tilde b x_j(t   - \tauc) 
e^{i2\pi (f_c + \nuc) \left( t -  \tauc \right)}
e^{i2\pi  
\beta  (j  + r N_T )   
}.
\]
Here, the approximation follows by the Doppler shift $\nuc$ being much smaller than the carrier frequency $f_c$, therefore $\frac{f_c + \nuc}{f_c} \approx 1$, and $\tilde \tau \approx \tauc$. 
If follows that the 
reflection of the $j$-th probing signal received by the $r$-th receive antenna, after demodulation, is 
\[
b x_j(t   - \tauc) 
e^{i2\pi \nuc t }
e^{i2\pi   \beta  (j  + r N_T ) }
\]
where we defined $b = \tilde b e^{-i2\pi \nuc \tauc}$. 
Next, consider $\S$ targets. 
Since, for $\S$ targets, the (demodulated) signal $y_r$ received by antenna $r$ consists of the superposition of the reflections of the probing signals $x_j,j=0,\ldots,N_T-1$, transmitted by the transmit antennas, 
we obtain the input-output relation \eqref{eq:iorelintro}, i.e.
\[
y_r(t)
=
\sum_{k=0}^{\S-1}
b_k 
e^{i2\pi r N_T \beta_k}
\sum_{j=0}^{N_T-1} 
 e^{i2\pi j \beta_k }  x_j(t - \tauc_k) e^{i2\pi \nuc_k t}.
\]
Here, $\beta_k, \tauc_k$ and $\nuc_k$ determine the angle, $\theta_k$, the distance,  $d_k$, and the velocity, $v_k$ of the $k$-th target relative to the antenna array.

\end{document}